\documentclass[twocolumn,english,prb,aps,english]{revtex4-1}
\usepackage[T1]{fontenc}
\usepackage[latin9]{inputenc}
\usepackage{float}
\usepackage{amsmath}
\usepackage{amssymb}
\usepackage{graphicx}
\usepackage{esint}

\makeatletter

\@ifundefined{textcolor}{}
{%
 \definecolor{BLACK}{gray}{0}
 \definecolor{WHITE}{gray}{1}
 \definecolor{RED}{rgb}{1,0,0}
 \definecolor{GREEN}{rgb}{0,1,0}
 \definecolor{BLUE}{rgb}{0,0,1}
 \definecolor{CYAN}{cmyk}{1,0,0,0}
 \definecolor{MAGENTA}{cmyk}{0,1,0,0}
 \definecolor{YELLOW}{cmyk}{0,0,1,0}
}

\@ifundefined{textcolor}{}{%
 \definecolor{BLACK}{gray}{0}
 \definecolor{WHITE}{gray}{1}
 \definecolor{RED}{rgb}{1,0,0}
 \definecolor{GREEN}{rgb}{0,1,0}
 \definecolor{BLUE}{rgb}{0,0,1}
 \definecolor{CYAN}{cmyk}{1,0,0,0}
 \definecolor{MAGENTA}{cmyk}{0,1,0,0}
 \definecolor{YELLOW}{cmyk}{0,0,1,0}
}

\usepackage{dcolumn}
\usepackage{bm}
\usepackage{enumerate}

\usepackage{babel}

\usepackage{babel}

\makeatother

\usepackage{babel}
\begin{document}

\title{Sublinear scaling for time-dependent stochastic density functional
theory}

\author{Yi Gao and Daniel Neuhauser}

\affiliation{Department of Chemistry and Biochemistry, University of California,
Los Angeles, CA-90095 USA}

\author{Roi Baer}

\affiliation{Fritz Haber Center for Molecular Dynamics, Institute of Chemistry,
The Hebrew University of Jerusalem, Jerusalem 91904, Israel}

\author{Eran Rabani}

\affiliation{Department of Chemistry, University of California and Lawrence Berkeley
National Laboratory, Berkeley, California 94720, USA}
\begin{abstract}
A stochastic approach to time-dependent density functional theory
(TDDFT) is developed for computing the absorption cross section and
the random phase approximation (RPA) correlation energy. The core
idea of the approach involves time-propagation of a small set of stochastic
orbitals which are first projected on the occupied space and then
propagated in time according to the time-dependent Kohn-Sham equations.
The evolving electron density is exactly represented when the number
of random orbitals is infinite, but even a small number ($\approx16$)
of such orbitals is enough to obtain meaningful results for absorption
spectrum and the RPA correlation energy per electron. We implement
the approach for silicon nanocrystals (NCs) using real-space grids
and find that the overall scaling of the algorithm is sublinear with
computational time and memory. 
\end{abstract}
\maketitle

\section{Introduction\label{sec:Introduction}}

Time-dependent density functional theory (TDDFT)~\cite{Runge1984}
allows for practical calculations of the time evolution of electronic
densities under time-dependent perturbations. In principle TDDFT is
an exact theory,\cite{Runge1984} but in applications, several assumptions
and approximations are typically made.\cite{Leeuwen2001,Onida2002,Marques2004,Burke2005a,Botti2007}
For example, for the most common usage of TDDFT, namely the absorption
spectrum~\cite{Jacquemin2009} of molecules and materials~\cite{Casida2009}
one uses the adiabatic approximation local/semi-local exchange-correlation
potentials (time-dependent adiabatic local density approximation\cite{Zangwill1980}
(TDALDA) or time-dependent adiabatic generalized gradient approximation
(TDAGGA)). TDDFT can also be used to compute the ground-state DFT
correlation energy within the adiabatic-connection fluctuation-dissipation
(ACFD) approach,\cite{Eshuis2012} or for studying strong-field nonpetrurbative
dynamics.\cite{Petersilka1999,Ceccherini2000,Chu2001a,Baer2003e,Nobusada2004,Castro2006}

There are two types of challenges facing the application of TDDFT
for large systems. One is the construction of appropriate functionals,
as the simplest, local and semi-local adiabatic functionals (TDALDA
and TDAGGA) often fail for large systems.\cite{Albrecht1998,Benedict1998,Rohlfing2000,Sottile2007,Ramos2008,Rocca2012a}
The second issue is the development of a linear-scaling approach that
overcomes not only the quartic ($O\left(N^{4}\right)$) scaling in
the frequency-domain formulation~\cite{Casida1996} but also the
quadratic ($O\left(N^{2}\right)$) limit achieved when real-time propagation
according to the time-dependent Kohn-Sham (TDKS) equations is used.\cite{Yabana1996,Bertsch2000,Baer2004b}
This latter scaling is commonly considered the lowest theoretical
scaling limit as it does not require full resolution of the TDKS excitation
energies. This is important for large systems where the density of
excited states is very large and there is no point in resolving of
all single-excited states as in small systems.

The present paper addresses the second challenge described above and
presents a stochastic formulation of TDDFT (TDsDFT) formally equivalent
to the TDKS method but \emph{without }the Kohn-Sham (KS) orbitals.
The new method is based on representing the time-dependent density
as an average over densities produced by \textit{evolving} projected
stochastic orbitals.\cite{Baer2013,Neuhauser2014} We consider two
demonstrations of the TDsDFT within the linear response limit: The
first concerns the calculation of the dipole absorption cross section
and the second is based on the ACFD approach to calculate the random
phase approximation DFT correlation energy . 

The paper is organized as follows: Section~\ref{sec:TD-sDFT} first
reviews the relation between linear response TDDFT and the generalized
susceptibility operator $\hat{\chi}^{\lambda}\left(t\right)$. Next,
we show how TDsDFT can be used to perform the time consuming computational
step in linear response applications, i.e. the action of $\hat{\chi}^{\lambda}\left(t\right)$
on a given potential. In Section~\ref{sec:results} we show how the
absorption spectrum and the ACFD-RPA correlation energy can be calculated
using TDsDFT. We present results for a series of silicon NCs of varying
sizes. We also analyze the scaling, accuracy and stability of the
proposed TDsDFT. In Section~\ref{sec:summary} we conclude.

\section{Theory\label{sec:TD-sDFT}}

\subsection{The Generalized Susceptibility Function and Time-Dependent Density
Functional Theory}

Consider a system of $N_{e}$ electrons interacting via a damped Coulomb
potential $(\lambda v_{C}\mathbf{\left(\left|\mathbf{r-}r'\right|\right)}$
where $0\le\lambda\le1$ and $v_{C}\left(r\right)=e^{2}/4\pi\epsilon_{0}r$
in their ground state $\left|0_{\lambda}\right\rangle $ and having
a density $n_{0}\left(\mathbf{r}\right)=\left\langle 0_{\lambda}\left|\hat{n}\left(\mathbf{r}\right)\right|0_{\lambda}\right\rangle .$
The linear density response of the system at time $t$ ($\delta n^{\lambda}\left(\mathbf{r},t\right)$)
to a small external time-dependent potential perturbation ($v\left(\mathbf{r}',t'\right)$)
is described by the following integral:\cite{Giuliani2005}

\begin{equation}
\delta n^{\lambda}\left(\mathbf{r},t\right)=\int_{0}^{t}dt'\int d\mathbf{r}'\chi^{\lambda}\left({\bf r},{\bf r}',t-t'\right)\delta v\left(\mathbf{r}',t'\right),\label{eq:linear-response}
\end{equation}
where $\chi^{\lambda}\left({\bf r},{\bf r}',t\right)$ is the generalized
susceptibility function,\cite{Fetter1971} which is also given by
retarded density-density correlation function of the system: 
\begin{equation}
\chi^{\lambda}\left({\bf r},{\bf r}',t\right)=\left(i\hbar\right)^{-1}\theta\left(t\right)\left\langle 0_{\lambda}\left|\left[\hat{n}^{\lambda}\left({\bf r},t\right),\hat{n}^{\lambda}\left({\bf r}',0\right)\right]\right|0_{\lambda}\right\rangle ,\label{eq:dens-dens}
\end{equation}
where $\hat{n}^{\lambda}\left({\bf r},t\right)$ is the density operator
at position $\mathbf{r}$ and time $t$. Eq.~\eqref{eq:dens-dens}
is also known as the fluctuation-dissipation relation.\cite{Kubo1995}
$\chi^{\lambda}\left({\bf r},{\bf r}',t\right)$ is used, for example,
to compute the linear polarizability and energy absorption of the
system under external fields, the dielectric response, the conductivity
and the correlation energies. 

Rather than computing $\chi^{\lambda}\left({\bf r},{\bf r}',t\right)$
directly (which in practice requires a huge effort for large systems),
a more efficient approach is to obtain $\delta n^{\lambda}\left(\mathbf{r},t\right)$
by applying an impulsive perturbation, i.e., $\delta v\left(\mathbf{r}',t'\right)=\gamma v\left(\mathbf{r}'\right)\delta\left(t'\right)$
($\gamma$ is a small constant with units of time):

\begin{equation}
\delta n^{\lambda}\left(\mathbf{r},t\right)=\gamma\int d\mathbf{r}'\chi^{\lambda}\left({\bf r},{\bf r}',t\right)v\left(\mathbf{r}'\right).\label{eq:linear-response-1}
\end{equation}
Here, $\delta n^{\lambda}\left(\mathbf{r},t\right)$ can be computed
by applying a perturbation $e^{-i\gamma\hat{v/\hbar}}$ and propagating
the perturbed ground state: 
\begin{equation}
\delta n_{\gamma}^{\lambda}\left({\bf r},t\right)=\left\langle 0_{\lambda}\left|e^{i\gamma\hat{v}/\hbar}\hat{n}^{\lambda}\left({\bf r},t\right)e^{-i\gamma\hat{v}/\hbar}\right|0_{\lambda}\right\rangle -n_{0}\left(\mathbf{r}\right),\label{eq:delta-n}
\end{equation}
where $\hat{v}=\int\hat{n}\left({\bf r}'\right)v\left({\bf r}'\right)d{\bf r}'.$
To see this, expand the right hand side of Eq.~\eqref{eq:delta-n}
to first order in $\gamma$: $\delta n_{\gamma}^{\lambda}\left({\bf r},t\right)=i\hbar^{-1}\gamma\left\langle 0_{\lambda}\left|\left[\hat{v},\hat{n}^{\lambda}\left({\bf r},t\right)\right]\right|0_{\lambda}\right\rangle $
which, when combined with \eqref{eq:dens-dens} gives Eq.~\eqref{eq:linear-response}. 

To obtain the density response $\delta n_{\gamma}^{\lambda}\left({\bf r},t\right)$
one needs to solve the many-electron time-dependent Schrödinger equation,
which is prohibitive in general. A practical alternative is to use
TDDFT. Starting from the KS system of non-interacting electrons having
the ground-state density $n_{0}\left(\mathbf{r}\right)=2\sum_{j\in occ}\left|\phi_{j}\left(\mathbf{r}\right)\right|^{2}$,
one perturbs the KS eigenstates $\phi_{j}\left(\mathbf{r}\right)$
at $t=0$:

\begin{equation}
\varphi_{j}\left({\bf r},t=0\right)=e^{-i\gamma v\left({\bf r}\right)/\hbar}\phi_{j}\left({\bf r}\right),\label{eq:phi t=00003D0-1}
\end{equation}
and then propagates in time according to the TDKS equations

\begin{equation}
i\hbar\frac{\partial\varphi_{j}\left({\bf r},t\right)}{\partial t}=\hat{h}^{\lambda}\left(t\right)\varphi_{j}\left({\bf r},t\right),\label{eq:td phi-2}
\end{equation}
where the TDKS Hamiltonian $\hat{h}^{\lambda}\left(t\right)$ depends
on the screening parameter $\lambda$ and the propagated density,
$n_{\gamma}^{\lambda}\left({\bf r},t\right)=2\sum_{j\in occ}\left|\varphi_{j}\left(\mathbf{r},t\right)\right|^{2}$.
The density response of Eq.~\eqref{eq:delta-n} is then obtained
from: 

\begin{align}
\int d\mathbf{r}'\chi^{\lambda}\left({\bf r},{\bf r}',t\right)v\left(\mathbf{r}'\right)= & \frac{1}{\gamma}\left(n_{\gamma}^{\lambda}\left({\bf r},t\right)-n_{\gamma=0}^{\lambda}\left({\bf r},t\right)\right)\nonumber \\
\equiv & \Delta n^{\lambda}\left(\mathbf{r,}t\right).\label{eq:delta n-t}
\end{align}
Eq.~\eqref{eq:delta n-t} simply states that the integral of the
susceptibility and a potential $v\left(\mathbf{r}\right)$ can be
computed from the difference between the perturbed and unperturbed
densities. This relation holds also for the half Fourier transform
quantities ($\tilde{f}\left(\omega\right)=\int_{0}^{\infty}dt\, e^{i\omega t}f\left(t\right)$):
\begin{align}
\int d\mathbf{r}'\tilde{\chi}^{\lambda}\left({\bf r},{\bf r}',\omega\right)v\left(\mathbf{r}'\right)= & \frac{1}{\gamma}\left(\tilde{n}_{\gamma}^{\lambda}\left({\bf r},\omega\right)-\tilde{n}_{\gamma=0}^{\lambda}\left({\bf r},\omega\right)\right)\nonumber \\
\equiv & \Delta\tilde{n}^{\lambda}\left(\mathbf{r,\omega}\right).\label{eq:delta n-omega}
\end{align}

\subsection{Time-Dependent \textit{Stochastic} Density Functional Theory}

The stochastic formulation of the density response is identical to
the deterministic version outlined above but instead of representing
the time-dependent density $n_{\gamma}^{\lambda}\left(\mathbf{r},t\right)$
as a sum over all occupied orbital densities ($|\varphi_{j}\left({\bf r},t\right)|^{2}$)
we represent it as an average over the densities of stochastic orbitals
$\xi_{j}\left(\mathbf{r},t\right)$.\cite{Baer2013} Each stochastic
orbital is first projected onto the occupied space and then propagated
in time. The advantage of the proposed approach is immediately clear:
If the number of stochastic orbitals needed to converge the results
does not increase with the system size $N$, the scaling of the approach
is linear with $N$ (rather than quadratic for the deterministic version).
Perhaps, in certain cases, due to self-averaging,\cite{Baer2013}
the scaling will even be better than linear, since the number of stochastic
orbitals required to converge the results to a predefined tolerance
may decrease with the system size.

The stochastic TDDFT (TDsDFT) procedure is outlined as follows (for
simplicity we use a real-space grid representation, but the approach
can be generalized to plane-waves or other basis sets):
\begin{enumerate}
\item Generate $N_{\zeta}$ stochastic orbitals $\zeta_{j}\left({\bf r}\right)=e^{i\theta_{j}\left({\bf r}\right)}/\sqrt{\delta V}$,
where $\theta_{j}\left({\bf r}\right)$ is a uniform random variable
in the range $\left[0,2\pi\right]$, $\delta V$ is the volume element
of the grid, and $j=1\,,\dots,\, N_{\zeta}$. Here, $N_{\zeta}$ is
typically much smaller than the number of total occupied orbitals
(more details below). The stochastic orbitals obey the relation $\mathbf{1}=\left<\left|\zeta\left\rangle \right\langle \zeta\right|\right>_{\zeta}$
where $\left<\cdots\right>_{\zeta}$ denotes a statistical average
over $\zeta$.
\item Project each stochastic orbital $\zeta_{j}\left({\bf r}\right)$ onto
the occupied space: $\left|\xi_{j}\right\rangle \equiv\sqrt{\hat{\theta}_{\beta}}\left|\zeta_{j}\right\rangle $,
where $\theta_{\beta}\left(x\right)=\frac{1}{2}\mbox{erfc}\left(\beta\left(\mu-x\right)\right)$
is a smooth representation of the Heaviside step function~\cite{Baer2013}
and $\mu$ is the chemical potential. The action of $\sqrt{\hat{\theta}_{\beta}}$
is performed using a suitable expansion in terms of Chebyshev polynomials~\cite{Kosloff1988}
in the static Hamiltonian with coefficients that depend on $\mu$
and $\beta$. 
\item As in the deterministic case, apply a perturbation at $t=0$: $\xi_{j}\left({\bf r},t=0\right)=e^{-i\gamma v\left({\bf r}\right)/\hbar}\xi_{j}\left({\bf r}\right)$
and propagate the orbitals according to the adiabatic stochastic TDKS
equations: 
\begin{align}
i\hbar\frac{\partial\xi_{j}\left({\bf r},t\right)}{\partial t} & =\hat{h}^{\lambda}\left(t\right)\xi_{j}\left({\bf r},t\right),\label{eq:tdsdft}
\end{align}
with $\hat{h}^{\lambda}\left(t\right)=\hat{h}_{KS}+v_{HXC}^{\lambda}\left[n_{\gamma}^{\lambda}\left(t\right)\right]\left(\mathbf{r}\right)-v_{HXC}^{\lambda}\left[n_{\gamma}^{\lambda}\left(0\right)\right]\left(\mathbf{r}\right)$
and 
\begin{equation}
v_{HXC}^{\lambda}\left[n\right]\left(\mathbf{r}\right)=\lambda\int d\mathbf{r}'\frac{n\left(\mathbf{r}'\right)}{\left|\mathbf{r}-\mathbf{r}'\right|}+v_{XC}^{\lambda}\left(n\left(\mathbf{r}\right)\right),
\end{equation}
where $v_{XC}^{\lambda}\left(n\left(\mathbf{r}\right)\right)$ is
the local density (or semi-local) approximation for the exchange correlation
potential. For convergence reasons $\hat{h}_{KS}$ is obtained with
a rather large number of stochastic orbitals using the sDFT~\cite{Baer2013}
(or its more efficient version, embedded fragment sDFT)\cite{Neuhauser2014}
and is fixed for the entire propagation. The difference term $v_{HXC}^{\lambda}\left[n_{\gamma}^{\lambda}\left(t\right)\right]\left(\mathbf{r}\right)-v_{HXC}^{\lambda}\left[n_{\gamma}^{\lambda}\left(0\right)\right]\left(\mathbf{r}\right)$
is generated with a relatively small number of stochastic orbitals
$N_{\zeta}$ and the density 
\begin{equation}
n_{\gamma}^{\lambda}\left({\bf {\bf r}},t\right)=2\left<\left|\xi\left({\bf r},t\right)\right|^{2}\right>_{\zeta}\approx\frac{2}{N_{\zeta}}\sum_{j=1}^{N_{\zeta}}\left|\xi_{j}\left({\bf r},t\right)\right|^{2}\label{eq:dens-from-xi}
\end{equation}
 is obtained as an \emph{average }over the stochastic orbital densities. 
\item Generate $\Delta n^{\lambda}\left({\bf r},t\right)=\frac{1}{\gamma}\left(n_{\gamma}^{\lambda}\left({\bf r},t\right)-n_{\gamma=0}^{\lambda}\left({\bf r},t\right)\right)$,
where $\gamma$ is a small parameter, typically $10^{-3}-10^{-5}\hbar E_{h}^{-1}$.
We note in passing that for $n_{\gamma=0}^{\lambda}\left({\bf r},t\right)$
one has to carry out the full propagation since the unperturbed projected
stochastic orbitals ($\left|\xi_{j}\right\rangle $) are not eigenstates
of the ground-state Hamiltonian. This propagation is not necessary
for the deterministic case. 
\end{enumerate}

\section{Results\label{sec:results}}

\subsection{TDsDFT Calculation of the Absorption Cross Section}

\begin{figure}[t]
\begin{centering}
\includegraphics[width=8cm]{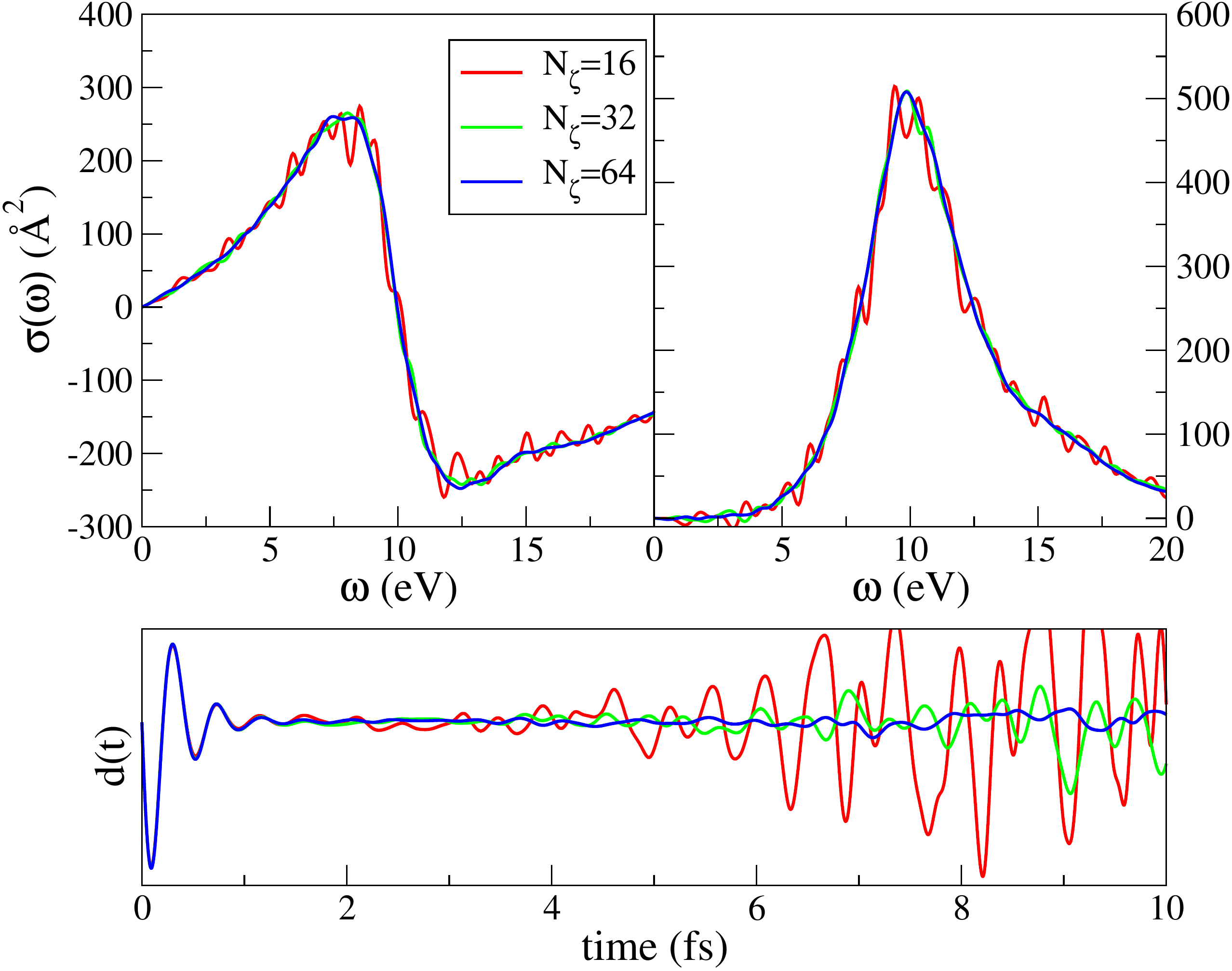}
\par\end{centering}

\protect\caption{\label{fig:absorption-cross-section} Upper panels: The real (left)
and imaginary (right) parts of $\sigma\left(\omega\right)$ calculated
for $\mbox{S\ensuremath{i_{705}}}\mbox{H}_{300}$ using $N_{\zeta}=16$~(red
line), $32$~(green line), and $64$~(blue line). Lower panel shows
the corresponding dipole correlation $d\left(t\right)$ as a function
of time.}
\end{figure}

The absorption cross section ($\omega\ge0$) is given by the imaginary
part of

\begin{equation}
\sigma\left(\omega\right)=\frac{e^{2}}{3\epsilon_{0}c}\omega\int d{\bf r}d{\bf r}'{\bf \, r}\cdot{\bf \tilde{\chi}\left({\bf r},{\bf r}',\omega\right)\cdot r}'.\label{eq:epsilon of omega}
\end{equation}
where $c$ is the speed of light. For simplicity, we assume that the
perturbing potential is in the $z\mbox{-direction}$ ($v\left({\bf r}\right)=z$)
and obtain $\sigma\left(\omega\right)$ in Eq.~\eqref{eq:epsilon of omega}
from the Fourier transform of the dipole-dipole correlation function:
\begin{equation}
d_{zz}\left(t\right)=\int z\Delta n_{z}^{\lambda=1}\left(\mathbf{r},t\right)d^{3}r,
\end{equation}
where $\Delta n_{z}^{\lambda=1}\left({\bf r},t\right)$ is obtained
from Eq.~\eqref{eq:delta n-t} and $\sigma\left(\omega\right)=\frac{e^{2}}{\epsilon_{0}c}\omega\int_{0}^{\infty}dt\, e^{i\omega t}d_{zz}\left(t\right)$. 

The real and imaginary parts of $\sigma\left(\omega\right)$ for $\mbox{S\ensuremath{i_{705}}}\mbox{H}_{300}$
are plotted in the upper panels of Fig.~\ref{fig:absorption-cross-section}.
These and all other results shown in this subsection were generated
using the algorithm above within the TDALDA approximation and a grid
representation with grid spacing of $\delta x=0.6a_{0}$ employing
norm-conserving pseudopotentials \cite{Troullier1991} and image screening
methods.\cite{Martyna1999} We used $\beta=0.01E_{h}^{-1}$ to represent
the smoothed step-function $\hat{\theta}_{\beta}$, and a Chebyshev
expansion length of $3770$ terms. The time-dependent dipole correlation
was calculated using a time step of $\delta t=0.0012\mbox{fs}$ up
to $t_{max}=7.5\mbox{fs}$. This signal was multiplied by a Gaussian
window function of width $2.5\mbox{fs}$ and then Fourier transformed
to give the absorption cross section.

The right upper panel of Fig.~\ref{fig:absorption-cross-section}
shows the absorption cross-section with a characteristic plasmon frequency
of $\sim10{\rm eV}$.\cite{Chelikowsky2003,Tsolakidis2005,Tiago2006,Ramos2008}
This feature is already captured with $N_{\zeta}=16$ stochastic orbitals
compared to $1560$ occupied orbitals required in the full deterministic
TDDFT. It is seen that further increase of $N_{\zeta}$ reduces the
statistical fluctuations and provides a handle on the accuracy of
the calculation. The convergence of the real part of $\sigma\left(\omega\right)$
shown in the upper left panel is similar to its imaginary counterpart.

The calculated dipole correlation $d_{zz}\left(t\right)$ is shown
in the lower panel of Fig.~\ref{fig:absorption-cross-section}. For
these large but finite systems we expect $d_{zz}\left(t\right)$ to
oscillate and decay to zero at intermediate times followed by recurrences
that appear at very long times (much longer than the timescales shown
here). Indeed, the stochastic approximation to $d_{zz}\left(t\right)$
oscillates and decays to zero up to a time $\tau_{C}$, but this is
followed by a gradual increase which eventually leads to divergence.
This is caused by an instability of the non-linear TDsDFT equations
due to the stochastic representation of the time-dependent density.
As $N_{\zeta}$ increases and the statistical fluctuations in the
density decrease, the divergence onset time $\tau_{C}$ is increased. 

\begin{figure}[H]
\begin{centering}
\includegraphics[width=8cm]{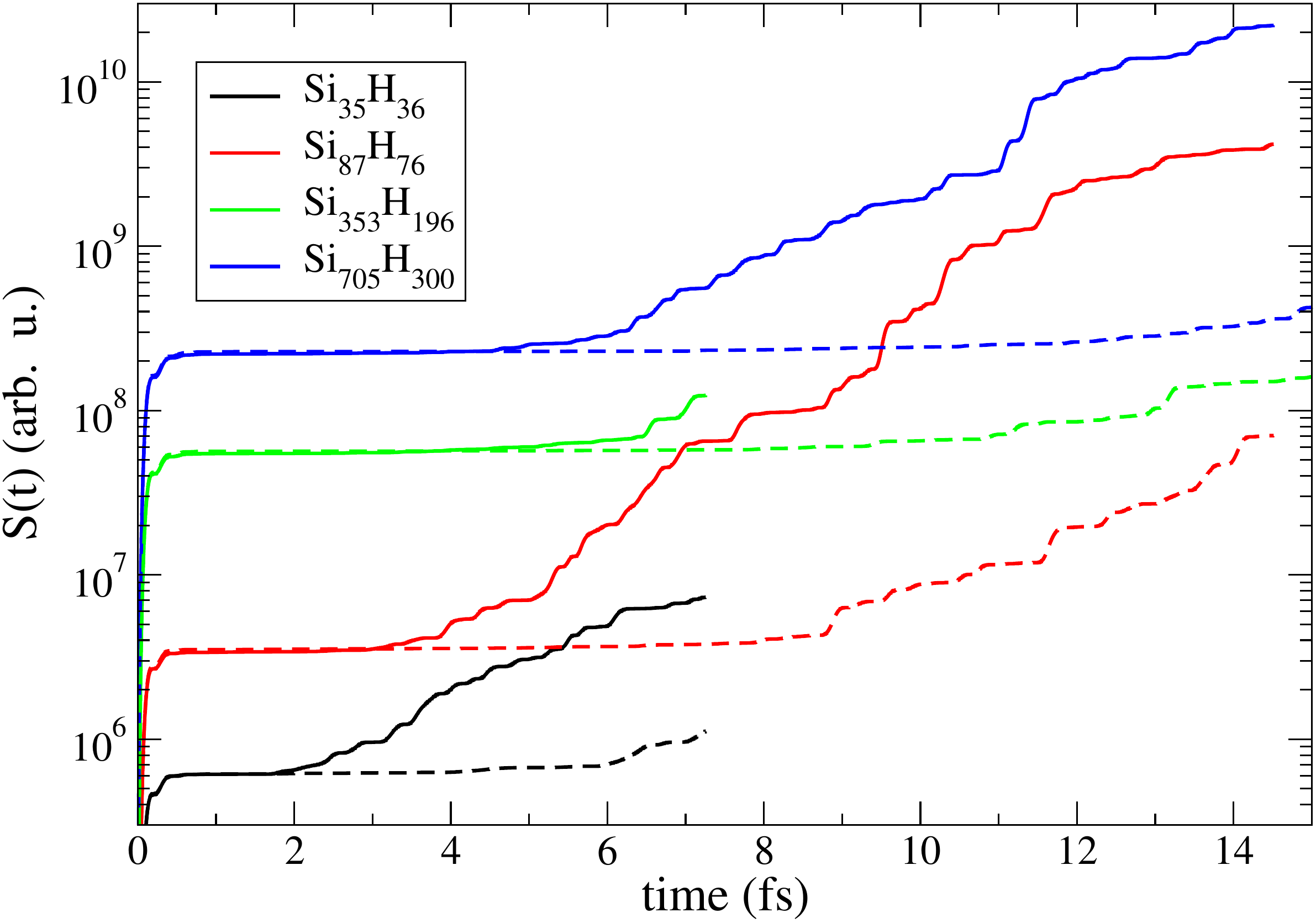}
\par\end{centering}

\protect\caption{\label{fig:divergence}Divergence of the stochastic TDDFT calculation.
Shown, the integrated dipole signal $S(t)=\int_{0}^{t}d_{zz}\left(t'\right)^{2}dt'$,
where $d_{zz}\left(t\right)$ is the calculated dipole correlation
as a function of time $t$, for $N_{\zeta}=16$ (solid line) and $N_{\zeta}=64$
(dashed line) stochastic orbitals for the different Si NCs. Because
of the decay of the dipole correlation the signals reach a plateau
after which they diverge sharply due to a nonlinear instability.}
\end{figure}
In Fig. \ref{fig:divergence} we plot the integrated dipole signal
$S(t)=\int_{0}^{t}d_{zz}\left(t'\right)^{2}dt'$ on a semi-log scale.
$S(t)$ provides a clearer measure of $\tau_{C}$, which is determined
as the onset of exponential divergence from the plateau (in practice
we take the value of $\tau_{C}$ to be at the middle of the plateau).
Two important observations on the onset of the divergence can be noted: 
\begin{enumerate}
\item $\tau_{C}$ increases for a fixed $N_{\zeta}$ as the system size
grows. For $N_{\zeta}=16$, $\tau_{C}$ increases from $\approx1.1\mbox{fs}$
for $\mbox{Si}_{35}\mbox{H}_{36}$ to $\approx2.3\mbox{fs}$ for $\mbox{Si}_{705}\mbox{H}_{300}$.
This is a rather moderate, but notable effect, that is a consequence
of the so called ``self-averaging''.\cite{Baer2013}
\item $\tau_{C}$ increases with $N_{\zeta}$ for a fixed system size. We
find that $\tau_{C}$ roughly scales as $N_{\zeta}^{1/2}$, namely,
an increase of $\tau_{C}$ by $2$ requires an increase of $N_{\zeta}$
by $4$. 
\end{enumerate}
These findings indicate that the number of stochastic orbitals not
only determines the level of statistical noise (which scales as $1/\sqrt{N_{\zeta}}$)
but also determines the spectral resolution, given by $\tau_{C}^{-1}$.
To achieve converged results for a fixed cutoff time of $\tau_{C}=10\mbox{fs},$
we find that $N_{\zeta}$ decreases from $\approx1300$ for $\mbox{Si}_{35}\mbox{H}_{36}$
to $\approx230$ for $\mbox{Si}_{705}\mbox{H}_{300}$.

\begin{figure}[H]
\begin{centering}
\includegraphics[width=8cm]{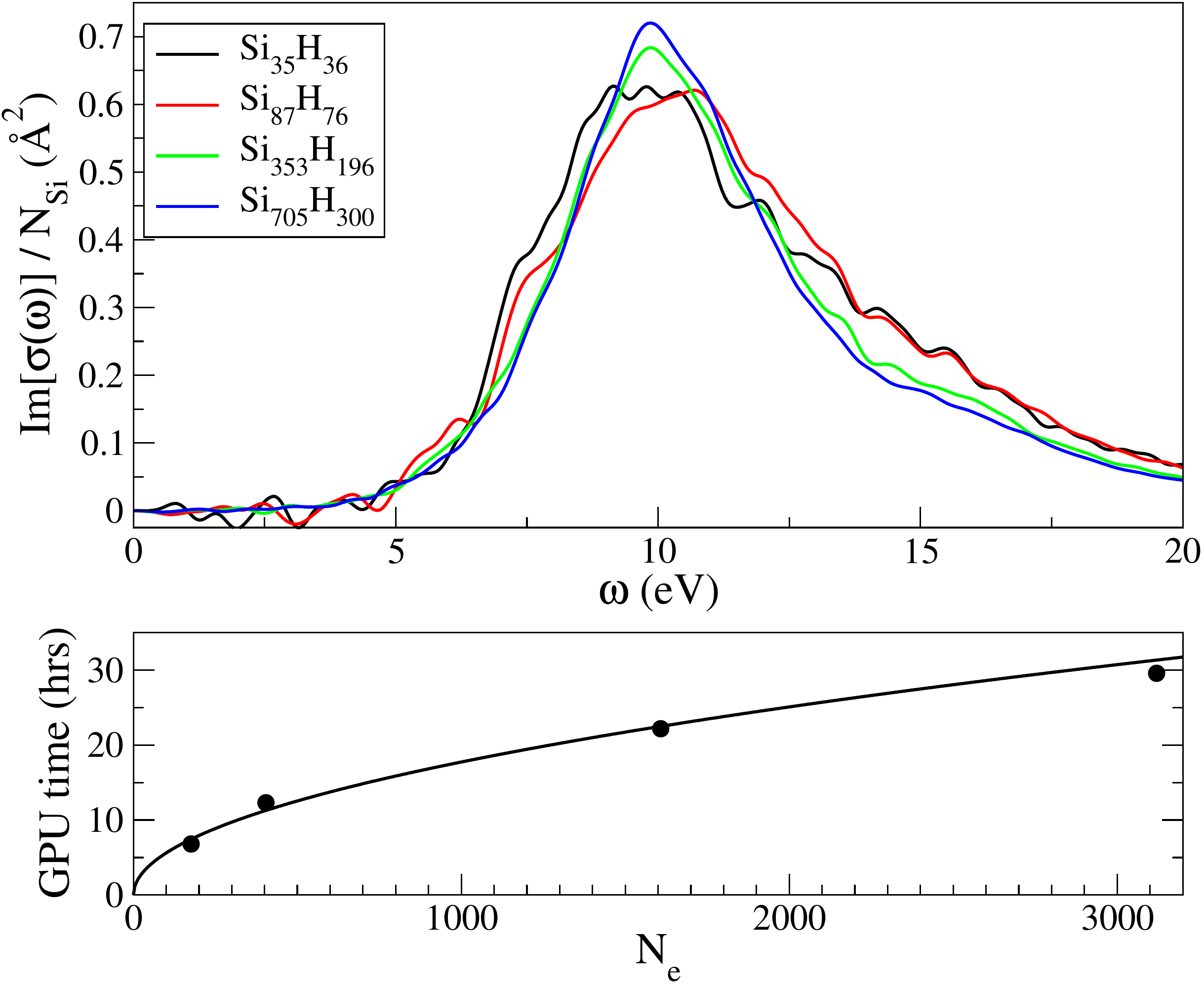}
\par\end{centering}

\protect\caption{\label{fig:Scaled-AbsCS}Upper panel: The absorption cross section
$\Im\sigma\left(\omega\right)$ scaled for size as calculated for
several silicon nanocrystals with $N_{\zeta}=64$. Lower panel: Extrapolated
GPU computational time ($T_{GPU}$) scaled for a cutoff time of $\tau_{C}=10\mbox{fs}.$
The line is a power law $T_{GPU}\propto N_{e}^{0.5}$.}
 
\end{figure}

In the upper panel of Fig.~\ref{fig:Scaled-AbsCS} we show the absorption
cross section for the series of silicon NCs and a fixed number of
stochastic orbitals, $N_{\zeta}=64$. As the NC size increases the
plasmon frequency (peak near $10\mbox{eV}$) slightly shifts to lower
energies and the width of the plasmon resonance slightly decreases.
This is consistent with classical Maxwell equations for which the
plasmon frequency depends strongly on the shape but very mildly on
the size of the NCs.\cite{Link1999} The statistical fluctuations
in the absorption cross section decrease with the system size for
a fixed $N_{\zeta}$, as clearly evident in the figure (most notably
at the lower energy range). 

The lower panel of Fig.~\ref{fig:Scaled-AbsCS} shows the GPU computational
time of the approach for a predefined spectral resolution (namely,
for converged results up to a fixed cutoff time $\tau_{C}=10\mbox{fs}$).
Each GPU performs roughly as $3$ Intel 3.5GHZ \textit{i}7 third generation
quad-core CPUs. Since the number of stochastic orbitals required to
converge the results for a fixed time decreases with the system size,
the overall scaling of the TDsDFT is better than $O\left(N_{e}\right)$
for the range of sizes studied here, significantly improving the $O\left(N_{e}^{2}\right)$
scaling of the full deterministic TDDFT. The overall computational
effort does depend on the spectral resolution and thus, for small
systems or for very high resolution the computational effort of the
stochastic approach may exceed that of the full deterministic calculation
with all occupied states. But this is certainly not the case for the
larger set of NCs studied here, where the wide plasmon resonance dominates
the absorption cross section, and thus the spectral features are converged
for $\tau_{C}<7.5\mbox{fs}$.

\subsection{Stochastic Approach to the Random Phase Approximation Correlation
Energy in DFT}

\begin{figure}[t]
\begin{centering}
\includegraphics[width=8cm]{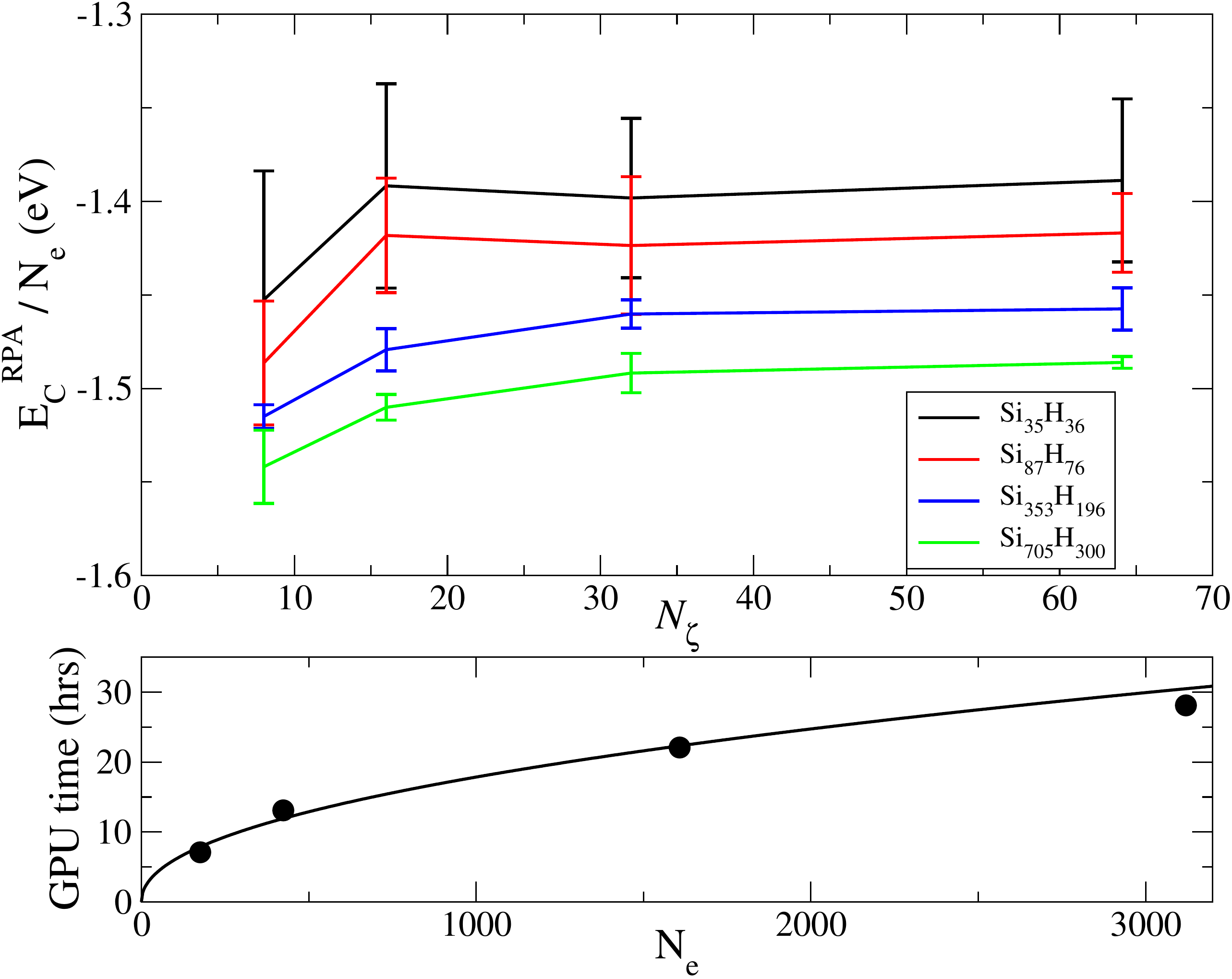}
\par\end{centering}

\protect\caption{\label{fig:The-RPA-Exc-vs-Nz}Upper panel: The RPA correlation energy
per electron for the silicon NCs as a function of the number of stochastic
orbitals $N_{\zeta}$ used to represent the time-dependent density.
The error bars are the standard deviation evaluated from 6 statistically
independent runs of the algorithm. Lower panel: GPU computational
time ($T_{GPU}$) scaled for a statistical error of $10\mbox{meV}$
in the total energy per electron. The line is a power law $T_{GPU}\propto N_{e}^{0.47}$.}
\end{figure}
The second application of the stochastic TDDFT is for the RPA correlation
energy, which is related to $\tilde{\chi}^{\lambda}\left({\bf r},{\bf r}',\omega\right)$
by the adiabatic-connection formula:\cite{Langreth1975,Gunnarsson1976}

\begin{align}
E_{C}^{RPA} & =-\frac{\hbar}{2\pi}\Im\int_{0}^{1}d\lambda\int_{0}^{\infty}d\omega\int d{\bf r}d{\bf r}'\times\nonumber \\
 & \left(\tilde{\chi}^{\lambda}\left({\bf r},{\bf r}',\omega\right)-\tilde{\chi}^{0}\left({\bf r},{\bf r}',\omega\right)\right)v_{C}\left(\left|{\bf r}-{\bf r}'\right|\right),\label{eq:E_CRPA}
\end{align}
where the integral over $\lambda$ adiabatically connects the non-interacting
density response $\tilde{\chi}^{0}\left({\bf r},{\bf r}',\omega\right)$
to the interacting one $\tilde{\chi}^{\lambda}\left({\bf r},{\bf r}',\omega\right)$.
To proceed, we rewrite Eq.~\eqref{eq:E_CRPA} as an average over
an additional set of stochastic orbitals $\eta\left({\bf r}\right)=e^{i\theta\left({\bf r}\right)}/\sqrt{\delta V}$ 

\begin{align}
E_{C}^{RPA} & =-\frac{\hbar}{2\pi}\Im\int_{0}^{1}d\lambda\int_{0}^{\infty}d\omega\int d{\bf r}d{\bf r}'d\mathbf{r}''\times\nonumber \\
 & \left\langle \eta^{*}\left({\bf r}\right)\left(\tilde{\chi}^{\lambda}\left({\bf r},{\bf r}',\omega\right)-\tilde{\chi}^{0}\left({\bf r},{\bf r}',\omega\right)\right)v_{C}\left({\bf r}'',{\bf r}'\right)\eta\left({\bf r}''\right)\right\rangle _{\eta}.\label{eq:E_CRPA-1}
\end{align}
This is done in order to rewrite the perturbation potential as a single-variable
potential: $v\left(\mathbf{r}'\right)=\int d{\bf r}'v_{C}\left({\bf r}',{\bf r}\right)\eta\left({\bf r}'\right)$,
which perturbs the stochastic orbitals at $t=0$: $\xi_{j}\left({\bf r},t=0\right)=e^{-i\gamma v\left({\bf r}\right)/\hbar}\xi_{j}\left({\bf r}\right)$
and from which the density $n_{\gamma}^{\lambda}\left({\bf {\bf r}},t\right)$
is computed using Eq.~\eqref{eq:dens-from-xi}. For the propagation
of $\xi_{j}\left({\bf r},t\right)$ according to Eq.~\eqref{eq:tdsdft}
we set $v_{XC}^{\lambda}\left(n\left(\mathbf{r}\right)\right)$ to
zero, i.e. use the time-dependent Hartree approximation. Using this
density and step 4 of the procedure outlined above we compute the
density response $\Delta n^{\lambda}\left({\bf r},t\right)$ from
which our the RPA correlation energy is calculated:

\begin{align}
E_{C}^{RPA} & =-\frac{\hbar}{2\pi}\Im\int_{0}^{1}d\lambda\int_{0}^{\infty}d\omega\int d{\bf r}\times\nonumber \\
 & \left\langle \eta^{*}\left({\bf r}\right)\left(\Delta\tilde{n}^{\lambda}\left({\bf r},\omega\right)-\Delta\tilde{n}^{\lambda=0}\left({\bf r},\omega\right)\right)\right\rangle _{\eta}.\label{eq:E_CRPA-1-2}
\end{align}
The stochastic formulation for Eq.~\eqref{eq:E_CRPA-1-2} follows
the algorithm described above in Sec.~\ref{sec:TD-sDFT}. 

We apply the stochastic RPA formulation to the various silicon NCs
studied above. The integration over $\lambda$ in Eq.~\eqref{eq:E_CRPA-1-2}
was carried out using Gaussian quadrature with $20$ sampling points.
For each value of $\lambda$ we used a different set of $\zeta$ (for
the TDsDFT) and $\eta$ (for the application of $v\left(\mathbf{r}\right)$)
stochastic orbitals. The TDsDFT total propagation time was $1.5\mbox{fs}$
with a time step $\delta t=0.0012\mbox{fs}$, sufficient to converge
the RPA correlation energy. 

In the upper panel of Fig.~\ref{fig:The-RPA-Exc-vs-Nz} we show the
calculated the RPA correlation energy per particle for the various
silicon NCs as a function of increasing $N_{\zeta}$, showing convergence
as $N_{\zeta}$ increases. The correlation energy per electron grows
(in absolute value) with system size, in accordance with our findings
in previous studies~\cite{Neuhauser2013,Neuhauser2013a} based on
a semi-empirical Hamiltonian~\cite{Wang1994d}. The standard deviation
(indicated by error bars) evaluated over $6$ different runs generally
decreases as $N_{\zeta}$ grows for a given system size and also decreases
as system size grows for a given value of $N_{\zeta}$. The magnitude
of the error, however, is rather noise due to the small number of
independent runs used to estimate it.

The lower panel of Fig.~\ref{fig:The-RPA-Exc-vs-Nz} shows the GPU
computational time of the approach for a fixed statistical error (estimated
as the standard deviation based on the estimate of 20 independent
runs) of $10\mbox{meV}$. Our previous stochastic formulation of the
RPA correlation energy relied on storing all occupied states (memory
wise scaled as $O\left(N_{e}^{2}\right)$) and the computational effort
of the RPA stage scaled as $O\left(N_{e}^{\alpha}\right)$ with $1<\alpha<2$,\cite{Neuhauser2013a}
better than quadratic scaling due to self-averaging. Comparing the
current approach with our previous work~\cite{Neuhauser2013a}, we
find that the present approach shows significant improvements with
respect to the computational time and memory requirements. The computational
time scales as $O\left(N_{e}^{0.47}\right)$ for the range of NCs
studied, better than linear scaling for the total RPA correlation
energy per electron.

\section{Summary\label{sec:summary}}

We have developed a stochastic approach to TDDFT for computing the
absorption cross section (via the time-dependent dipole correlation
function) and the RPA correlation energy. The core idea of the approach
involves time propagation of a set of $N_{\zeta}$ stochastic projected
orbitals $\xi_{j}\left(\mathbf{r},t\right)$ according to the time-dependent
Kohn-Sham equations. The evolving electron density is exactly represented
when $N_{\zeta}\to\infty$ but the strength of the method appears
when a small number of orbitals $N_{\zeta}\ll N_{e}$, where $N_{e}$
is the number of electrons, is used. Such a truncation produces a
statistical fluctuation due to finite sampling. The magnitude of this
error is proportional to $1/\sqrt{N_{\zeta}}$.

The finite sampling error coupled with a nonlinear instability of
the time-dependent Kohn-Sham equations produces a catastrophic exponential
divergence that becomes noticeable only after a certain propagation
time $\tau_{C}$, which determines the spectral resolution of the
approach. The onset of divergence can be controlled by increasing
$N_{\zeta}$ and empirically we determined that $\tau_{C}\propto\sqrt{N_{\zeta}}$,
consistent with the statistical nature of the error. 

The TDsDFT was applied to study the absorption cross section and RPA
correlation energy for a series of silicon NCs with sizes as large
as $N_{e}\approx3000$. For this range of NC sizes, the computational
time scales sub-linearly, roughly as $O\left(N_{e}^{1/2}\right)$
for both the absorption cross section and for the RPA correlation
energy per electron. For the former, the scaling holds for a given
spectral resolution $\tau_{C}$. Since the computational time is also
proportional to $N_{\zeta}N_{e}$, one can work backwards to show
that $\tau_{C}\propto\sqrt{N_{\zeta}N_{e}^{1/2}}$. For the RPA application,
the scaling holds for a given statistical error in the RPA correlation
energy per electron. 

The developed stochastic TDDFT approach adds another dimension to
the arsenal of stochastic electronic structure methods, such as the
sDFT~\cite{Baer2013} (and its more accurate fragmented version)\cite{Neuhauser2014}
and the sGW.\cite{Neuhauser2014a} Future work will extend the approach
to include exact and screened exchange potentials in order to account
for charge-transfer excited states and multiple excitations. 
\begin{acknowledgments}
RB and ER are supported by The Israel Science Foundation -- FIRST
Program (grant No. 1700/14). Y. G. and D. N. are part of the Molecularly
Engineered Energy Materials (MEEM), an Energy Frontier Research Center
funded by the DOE, Office of Science, Office of Basic Energy Sciences
under Award No. DE-SC0001342. D. N. also acknowledges support by the
National Science Foundation (NSF), Grant .CHE-1112500. 
\end{acknowledgments}

\end{document}